\documentclass[
groupedaddress,
 amsmath,amssymb,
 aps,
pra,
]{revtex4-2}

\usepackage{graphicx}
\usepackage{dcolumn}
\usepackage{bm}
\usepackage{xcolor}
\usepackage{colortbl}      

\begin{document}

\title[Light-induced, fictitious magnetic trapping of cold alkali atoms]{Light-induced, fictitious magnetic trapping of cold alkali atoms using an optical tweezers-nanofiber hybrid platform}

\author{Alexey Vylegzhanin}
\thanks{Corresponding author:~alexey.vylegzhanin@oist.jp}
\address{Light-Matter Interactions for Quantum Technologies Unit, Okinawa Institute of Science and Technology Graduate University, Onna, Okinawa 904-0495, Japan.}

\author{Dylan J. Brown}
\altaffiliation[Current address: ]{Centre for Cold Matter, Blackett Laboratory, Imperial College London, Prince Consort Road, London, SW7 2AZ UK}
\address{Light-Matter Interactions for Quantum Technologies Unit, Okinawa Institute of Science and Technology Graduate University, Onna, Okinawa 904-0495, Japan.}

\author{Sergei Abdrakhmanov}
\address{Light-Matter Interactions for Quantum Technologies Unit, Okinawa Institute of Science and Technology Graduate University, Onna, Okinawa 904-0495, Japan.}

\author{S\'ile Nic Chormaic}
\thanks{Corresponding author:~sile.nicchormaic@oist.jp}
\address{Light-Matter Interactions for Quantum Technologies Unit, Okinawa Institute of Science and Technology Graduate University, Onna, Okinawa 904-0495, Japan.}


\begin{abstract}
We present a magnetic trapping scheme for cold $^{87}$Rb atoms based on light-induced fictitious magnetic fields generated by the evanescent field of an optical nanofiber (ONF) integrated with an optical tweezers. We calculate and compare the trapping potentials for both  Gaussian and Laguerre-Gaussian modes of the  tweezers beam, combined with a quasi-linearly polarized ONF-guided field. Based on the optical powers in the tweezers and  ONF modes, we analyze the trap depths and the positions of the potential minima  from the nanofiber surface.  We show that, by varying the optical powers in the two fields, the trap position can be tuned over several hundred nanometers, while simultaneously influencing the trap depth and trap frequencies. Such control over atom-surface position is essential for studying distance-dependent effects on atoms trapped near a dielectric surface and optimizing atom-photon interfaces for quantum technology applications. 
\end{abstract}

\maketitle


\section{Introduction}
In recent years, there has been growing interest in trapping and optically interfacing cold, neutral atoms near optical waveguides~\cite{balykin2004atom, vetsch2010optical,thompson2013coupling,daly2014nanostructured,da2020cold,le2021optical, bouscal2024systematic,li2024atom,liu2025ferromagnetic,sadeghi2024long}. One highly efficient trapping  method uses optical dipole forces from  blue- and red-detuned evanescent light fields of an optical nanofiber (ONF), forming what is referred to as a two-color trap~\cite{balykin2004atom}. There have been several experimental realizations of such a scheme for laser-cooled Cs~\cite{vetsch2010optical,goban2012demonstration,white2019cavity} and Rb~\cite{Lee_2015, gupta2022machine} atoms. Furthermore, trapping schemes based on three-color configurations have also been proposed for trapping rubidium atoms via an ONF~\cite{vera2024nanofiber}. Aside from trapping, the evanescent field can  be used to probe atoms near the nanofiber ~\cite{sague2007cold,russell2011sub,lee2013integrated,kumar2015autler,kato2019observation,ray2020observation} or for the excitation of  atoms to Rydberg states~\cite{stourm2020spontaneous, KP_rydberg_generation, stourm2023interaction,  Vylegzhanin:23, zhang2025unidirectional}. The latter technique is promising for quantum computation or the realization of quantum networks/repeaters~\cite{li2024atom, wilk2010entanglement,ebert2015coherence, calajo2016atom, cong2022hardware, heller2022raman}.  

Another rapidly developing platform for trapping and manipulating  cold  atoms in well-controlled sites is that of free-space optical tweezers ~\cite{frese2000single,beugnon2007two,kaufman2012cooling,PhysRevA.102.063107,manetsch2025tweezer}.  Such systems have been extended to the trapping of Rydberg  atoms~\cite{barredo2020three,wilson2022trapping}.  The advantage of optical tweezers lies in their ability to trap a large number of atoms simultaneously in complex 1D, 2D, or 3D arrays, and their integration into cold atom setups is very well-established. 
In addition, optical tweezers can be combined with optical waveguides, such as optical nanofibers~\cite{nayak2019real} or photonic crystal waveguides~\cite{beguin2020reduced}, to trap atoms near the waveguide surface, providing enhanced coupling into the guided mode. However, such a scheme relies on the reflection of the  tweezers beam from the waveguide itself, leading to the atom being trapped at a fixed distance from the surface. The position of the trap minimum can be changed by varying the geometric parameters of the waveguide (which is impossible to do in situ), by tuning the tweezers  wavelength over hundreds of nanometers (only possible if using widely tunable lasers) or by changing the beam polarization~\cite{beguin2020reduced}, for which the fastest achievable time is on the order of $\mu$s~\cite{mitchell2016high}.

In this work, we propose a method to trap ground state, laser-cooled $^{87}$Rb atoms in a light-induced magnetic field trap~\cite{vylegzhanin2025towards}, formed by combining the evanescent field of an ONF with a circularly polarized optical tweezers beam consisting of either a Gaussian or  a Laguerre-Gaussian (LG) mode. We term the hybrid platform of both optical tools "OPTON" - OPtical Tweezers and Optical Nanofiber. We choose $^{87}$Rb since it is one of the more common elements for experiments with Rydberg states. Atoms in an electromagnetic field with nonzero ellipticity experience a light-induced fictitious magnetic field~\cite{cohen1972experimental,KienEPJD2013,schneeweiss2014nanofiber,Chen2025FictitiousBfield}. From now on, we will assume that this light-induced B-field is fictitious without explicitly writing this each time. In our proposed trapping scheme using OPTON, both the evanescent field and the optical tweezers have  nonzero ellipticity and, therefore, produce a light-induced magnetic field for the atoms. We set the polarization of the ONF-guided light and the optical tweezers  so that a local minimum of the fictitious magnetic field is formed. The distance between the ONF surface and the position of the trap minimum can be adjusted by changing the power of either the ONF-guided field or the optical tweezers  using acousto-optical modulators (AOMs) on timescales of $\mu$s comparable to current optical tweezers experiments for cold, neutral atoms~\cite{doi:10.1126/science.aah3778,PhysRevA.102.063107}. The proposed trapping method may be adaptable to create rings of atoms around optical nanofibers for studying fiber-mediated collective interactions~\cite{liedl2024observation, olmos2025hybrid, jimavsnez2025collective}.

\section{Light-induced, fictitious magnetic field}

An atom in an oscillating electromagnetic field experiences an AC-Stark shift, which depends on the strength of the electric field and the atom's frequency-dependent polarizability, thereby shifting the energy levels. This polarizability can be represented by the scalar, vector, and tensor components of the polarizability tensor~\cite{KienEPJD2013}. By using a tune-out wavelength, the scalar light shift can be minimized to near zero, leaving only the vector and tensor light shifts. For the $^{87}\mathrm{Rb}$ ground state, $\mathrm{5S}_{1/2}$, the tensor light shift is zero, and the tune-out wavelength is $790.2$~nm~\cite{leblanc2007species}. At this wavelength, the scalar light shifts from the D$_1$ and D$_2$ transitions cancel each other out in the $\mathrm{5S}_{1/2}$ state, resulting in a net zero shift. Consequently, the only remaining component is the vector light shift, which can be written as 

\begin{equation} \label{eq:1}
	\Delta E^\mathrm{v}_{\mathrm{AC}}=\frac{1}{4} \alpha^\mathrm{v}_{nJF}\: i [\boldsymbol{\mathcal{E}}^*\times \boldsymbol{\mathcal{E}}] \frac{m_F}{2F}.
\end{equation}

\noindent Here, $n$ is the principle quantum number, $J$ is the total angular momentum quantum number, $F$ is the hyperfine splitting quantum number, $m_F$ is the Zeeman splitting atomic level, $\alpha^{\mathrm{v}}_{nJF}$ is the frequency-dependent vector polarizability of an atom in the $\left|nJF\right>$ state, and $\boldsymbol{\mathcal{E}}$ is the positive-frequency electric field envelope of the complex electric field, $\mathbf{E}=1/2\left(\boldsymbol{\mathcal{E}}e^{-i\omega t} + \mathrm{c.c.}\right)$. The vector light shift depends on the magnetic quantum number, $m_F$, and can be interpreted as the magnetic potential ${U_\mathrm{mag}=\mu_{\mathrm{B}} g_{nJF} m_F |\boldsymbol{B}_\mathrm{fict}|}$, where $\boldsymbol{B}_\mathrm{fict}$ is a light-induced, fictitious magnetic field, expressed by~\cite{KienEPJD2013, schneeweiss2014nanofiber}

\begin{equation} \label{eq:2}
	\boldsymbol{B}_{\mathrm{fict}}=\frac{\alpha^{\mathrm{v}}_{nJF}}{8\mu_{\mathrm{B}} g_{nJF} F}\:i[\boldsymbol{\mathcal{E}}^*\times \boldsymbol{\mathcal{E}}],
\end{equation}
where, $\mu_{\mathrm{B}}$ is the Bohr magneton and  $g_{nJF}$ is the Landé g-factor.

\section{Light-induced, fictitious magnetic trap}
To model the fictitious magnetic field trap, we introduce an optical nanofiber made of silica with refractive index $n=1.44$ and radius $a=175$~nm. A fiber of this radius only supports the HE$_{11}$ mode of the 790.2~nm light used in this work~\cite{frawley2012higher,PhysRevA.96.023835}. A significant portion of the guided light extends outside the optical nanofiber in the form of an evanescent field~\cite{le2004field}. We set the polarization of the guided mode to quasi-linear along the $x$-axis in the ONF region, meaning that there are only $z$ and $x$ components of the electric field. 

\begin{figure*}
	\centering
	\includegraphics[width=\linewidth]{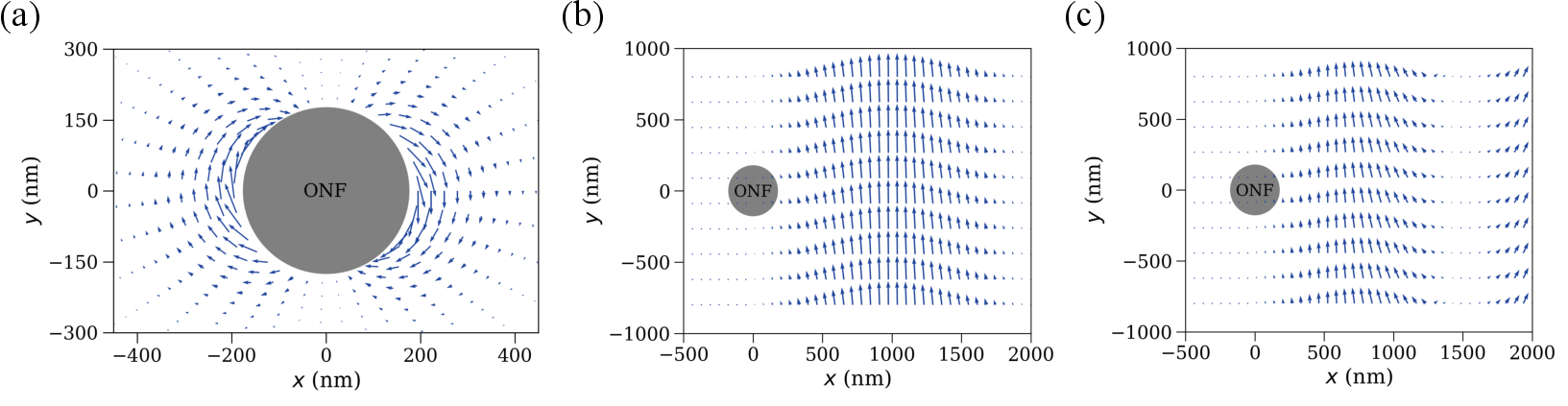}
	\caption{Vector field of the light-induced fictitious magnetic field (blue arrows) for an atom in the $\mathrm{\left|5P_{1/2}, F=2, m_F=2\right>}$ state in the $xy-$plane perpendicular to the fiber for (a) a quasi-linearly polarized along the $x$--axis fundamental ONF guided mode, (b) a circularly polarized Gaussian mode, and (c) an LG$_{01}$ mode tweezers. The fiber radius is $a=175$~nm and the free-space wavelength of the ONF guided mode is $\lambda_{\mathrm{ONF}}=787.9$~nm. The waist, $w_0$, is 500~nm, the distance between the center of the beam and the ONF surface is 825~nm (1325~nm) for the Gaussian (LG) mode, the power of the tweezers beam is 0.3~mW (regardless of mode chosen), and the free-space wavelength is $\lambda_{\mathrm{tw}}=790.2$~nm. One can see that the LG mode tweezers beam produces a larger light-induced magnetic field closer to the fiber surface. In addition, for both the LG and Gaussian modes, an $x-$component of the fictitious magnetic field is present due to the non-paraxial regime.}
	\label{fig:vector field}
\end{figure*}

The fictitious magnetic field  presented in Equation~\ref{eq:2}, created by an ONF guided mode, can be expressed in cylindrical coordinates as
\begin{equation}\label{eq:simpl_fict}
	\boldsymbol{B}_{\mathrm{fict}}=\frac{\alpha^{\mathrm{v}}_{nJF}}{4\mu_{\mathrm{B}} g_{nJF} F}[\mathrm{Im}(\mathcal{E}_z\mathcal{E}^*_r) \hat{\mathbf{\boldsymbol{\phi}}}+\mathrm{Im}(\mathcal{E}_r\mathcal{E}^*_\phi)\hat{\boldsymbol{z}}+\mathrm{Im}(\mathcal{E}_\phi\mathcal{E}^*_z)\hat{\boldsymbol{r}}],
\end{equation}
\noindent where $\boldsymbol{\mathcal{E}}_\mathrm{circ}^{(fp)}=(\mathcal{E}_r, \mathcal{E}_{\phi}, \mathcal{E}_z)$ represents the radial, azimuthal, and longitudinal cylindrical components of the electric field of the nanofiber-guided mode that are precisely described by Le Kien \textit{et al.}~\cite{PhysRevA.96.023835}. In the case of the quasi-linearly (QL) polarized guided mode of the ONF, the fictitious magnetic field can be simplified to the following form

\begin{equation}\label{eq:QL_B}
	\textbf{B}_{\mathrm{fict}}=\frac{\alpha^{\mathrm{v}}_{nJF}}{4\mu_{\mathrm{B}} g_{nJF} F}[\mathrm{Im}(\mathcal{E}_{z, \mathrm{lin}}\mathcal{E}^*_{r,\mathrm{lin}}) \hat{\boldsymbol{\phi}}+\mathrm{Im}(\mathcal{E}_{\phi, \mathrm{lin}}\mathcal{E}^*_{z,\mathrm{lin}}) \hat{\boldsymbol{r}}], 
\end{equation}

\noindent where $\boldsymbol{\mathcal{E_{\mathrm{lin}}}}$ is the electric field of a linearly polarized mode and is written as a summation of opposite handedness quasi-circular (QC)  polarization guided modes

\begin{equation}
	\boldsymbol{\mathcal{E}}_{\mathrm{lin}}^{(f\phi_\mathrm{pol})}=\frac{1}{\sqrt{2}}(\boldsymbol{\mathcal{E}}_\mathrm{circ}^{(f+)}e^{-i\phi_\mathrm{pol}}+\boldsymbol{\mathcal{E}}_\mathrm{circ}^{(f-)}e^{i\phi_\mathrm{pol}}).
\end{equation}

\noindent Here, $\phi_\mathrm{pol}$ is the polarization angle with respect to the $x-$axis, $f$ is either $+1$ or $-1$ and determines the propagation direction, with $+$ and $-$ defining the handedness of the QC guided mode~\cite{le2013state}. A vector plot of the light-induced magnetic field in the $xy$-plane from the QL guided mode of the ONF is shown in Fig.~\ref{fig:vector field}(a).

\begin{figure}[t]
    \centering
    \includegraphics[width=0.95\linewidth]{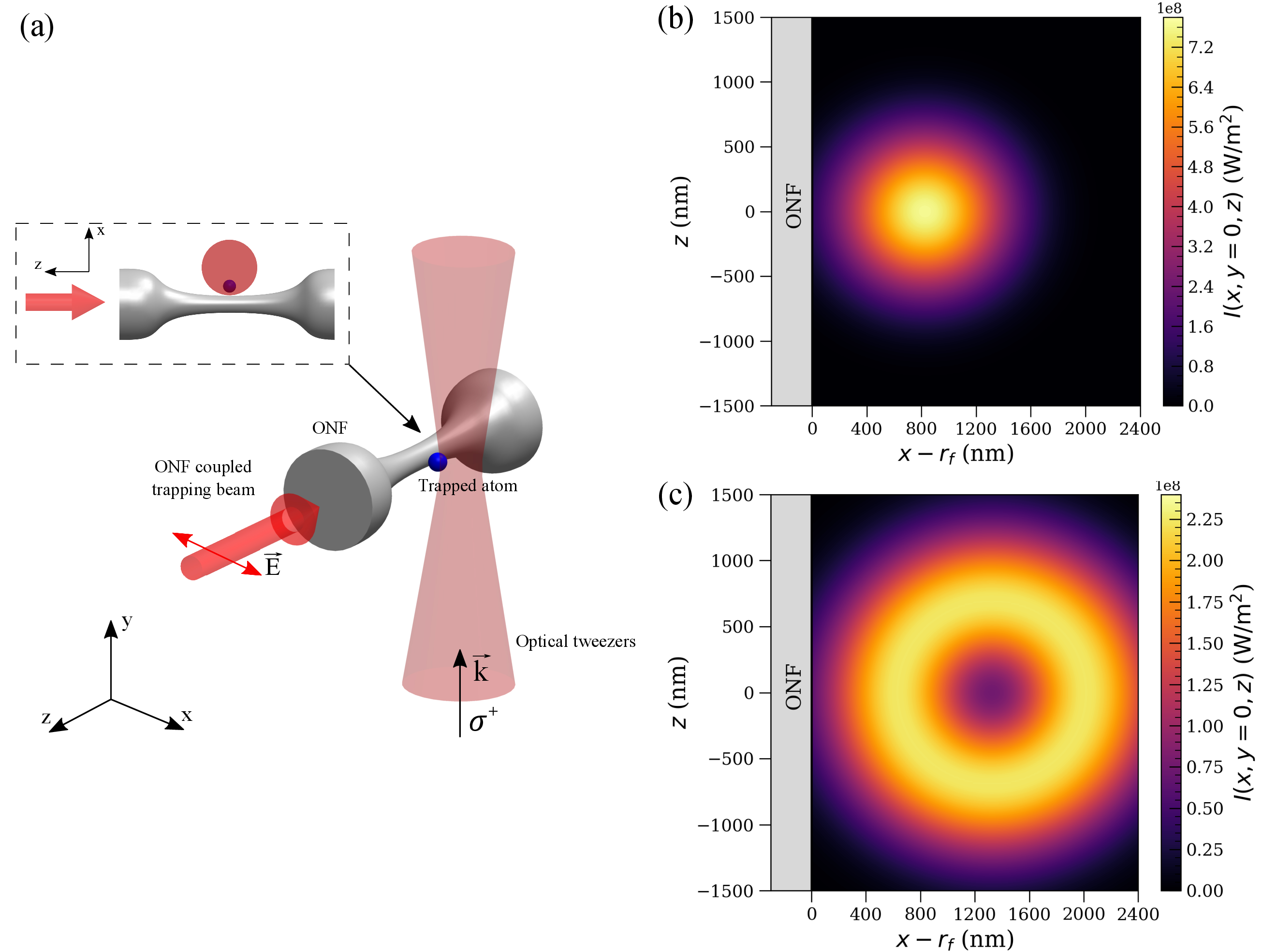}
    \caption{(a) Schematic of the experimental setup. A light beam (intense red arrow) with a free-space wavelength $\lambda_{\mathrm{ONF}}$ is coupled into the optical nanofiber (ONF), exciting the fundamental, quasi-linearly polarized guided mode ($\vec{E}$). An optical tweezers beam (faint red), of either a Gaussian or Laguerre--Gaussian mode, is focused at a distance $d = 825~\mathrm{nm}$ or $d = 1325~\mathrm{nm}$ from the fiber surface, respectively. The beam waist is $w_0 = 500~\mathrm{nm}$ and the tweezers wavelength is $\lambda_{\mathrm{tw}}$. A single atom (blue sphere) is trapped between the ONF surface and the tweezers focus beam due to the combined potentials of the evanescent field and the focused beam. The propagation direction and polarization of the tweezers are indicated by $\vec{k}$ and $\sigma^+$, respectively. The inset shows the $xz$-plane at $y = 0$. (b,c) Calculated intensity profiles $I(x,z)$ of the Gaussian and Laguerre--Gaussian modes with $0.3~\mathrm{mW}$ of power in the $xz$--plane at $y = 0$.}
    \label{fig:setup}
\end{figure}


The polarization of the optical tweezers beam is set to clockwise in the $xz-$plane while the Poynting vector is set along the $y-$direction. Therefore, the fictitious magnetic field is directed along the $y$-axis. We set the center of the optical tweezers at 825~nm from the surface of the ONF and the waist of the tweezers beam, $w_0$, i.e. the $1/e^2$ radius of the beam  spot size, to 500~nm. The power of the optical tweezers is set to 0.3~mW. The schematic of the setup with the OPTON arrangement is shown in Fig.~\ref{fig:setup}(a).

The electric field of the linearly polarized optical tweezers in the non-paraxial regime near the focal plane can be calculated using Richards–Wolf expressions~\cite{richards1959electromagnetic, born2013principles} and has the form~\cite{winnerl2012universal}

\begin{align}
\boldsymbol{E_x}& = [-iA \int_0^{\alpha} E_0(\theta) \, \sin \theta \left[ (1 + \cos \theta) J_0(k r_p \sin \theta \sin \theta_p) \right. \nonumber \\
& \qquad \left. + (1 - \cos \theta) J_2(k r_p \sin \theta \sin \theta_p) \cos 2\phi_p \right] e^{i k r_p \cos \theta \cos \theta_p} \, d\theta]\hat{e_x},
\end{align}

\begin{align}
\boldsymbol{E_{x}^y} &= [-2A \int_0^{\alpha} E_0(\theta) \, \sin^2 \theta \, \cos \phi_p \, J_1(k r_p \sin \theta \sin \theta_p) \, e^{i k r_p \cos \theta \cos \theta_p} \, d\theta]\hat{e_y},
\end{align}

\noindent where 

\begin{equation}
    \begin{split}
        &E_0(\theta) = \exp\left[-\beta_0^2 \left( \frac{\sin\theta}{\sin\alpha} \right)^2 \right]\sqrt{\cos{\theta}},
        \\        &\alpha=\arcsin(\mathrm{NA}).
    \end{split}
\end{equation}


\noindent Here, $\boldsymbol{E_x}$ and $\boldsymbol{E_{x}^y}$ are the transverse and longitudinal electric field components,  $\mathrm{NA}\approx0.5$, $\beta_0=3/2$ is the focusing parameter, $(r_p,\phi_p,\theta_p)$ are the spherical coordinates, $\theta$ is the angle under which the electric field from the lens is ``seen" at the focal point, and $A$ is a normalization factor. $J_{0,1,2}$ are the Bessel functions of the first kind and zeroth, first and second order, respectively. We can now construct the electric field of the circularly polarized optical tweezers from the electric field of the linearly polarized tweezers in the following way

\begin{equation}\label{eq:circ_fields}
    \begin{split}
        &\boldsymbol{E_{xz}=\frac{1}{\sqrt{2}}(E_x+iE_z)}
        \\
        &\boldsymbol{E_y=\frac{1}{\sqrt{2}}(E_{x}^y+E_{z}^y)}
    \end{split}
\end{equation}

\noindent Since $|E|^2=2\eta_0I$ and $I=\frac{2P}{\pi\omega_0^2}$, where $I$ is the intensity, $P$ is the power, $\eta_0=377~\Omega$ is the vacuum impedance, and $\omega_0$ is the waist of the focused beam, and, based on Equations~\ref{eq:simpl_fict} and \ref{eq:circ_fields}, the fictitious magnetic field can be written in the following form~\cite{corwin1999spin,yang2008optically,schneeweiss2014nanofiber}
\begin{equation}\label{eq:fict_field}
    \begin{split}
&\boldsymbol{B}_\mathrm{fict}^{\mathrm{tw}_y}=\frac{4\eta_0P}{\pi\omega_0^2}\frac{\alpha^{\mathrm{v}}_{nJF}}{4\mu_B g_{nJF} F}\mathrm{Im}(\boldsymbol{E_x}\cdot\boldsymbol{E_z^*})\boldsymbol{\hat{e_y}},
    \\
&\boldsymbol{B}_\mathrm{fict}^{\mathrm{tw}_z}=\frac{4\eta_0P}{\pi\omega_0^2}\frac{\alpha^{\mathrm{v}}_{nJF}}{4\mu_B g_{nJF} F}\mathrm{Im}(\boldsymbol{E_x^y}\cdot\boldsymbol{E_x}^*)\boldsymbol{\hat{e_z}},
\\
&\boldsymbol{B}_\mathrm{fict}^{\mathrm{tw}_x}=\frac{4\eta_0P}{\pi\omega_0^2}\frac{\alpha^{\mathrm{v}}_{nJF}}{4\mu_B g_{nJF} F}\mathrm{Im}(\boldsymbol{E_z}\cdot\boldsymbol{E_x^{y*}})\boldsymbol{\hat{e_x}}.
    \end{split}
\end{equation}

\begin{figure}
    \centering
    \includegraphics[width=1\linewidth]{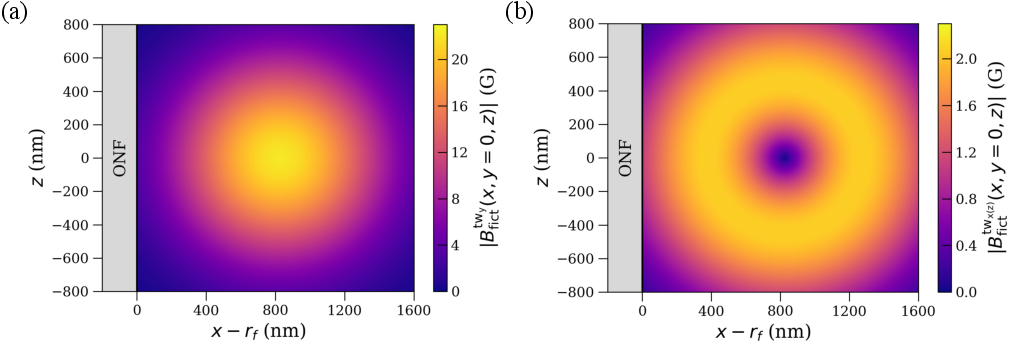}
    \caption{Two-dimensional plots of the amplitude of the  light-induced, fictitious magnetic field generated by a Gaussian optical tweezers beam with power  $P = 0.3~\mathrm{mW}$ in the nonparaxial regime for   (a) the $y$--component and (b) the $x(z)$--component. The fiber center is located at $x=0~( \mathrm{or}~(x-r_f)=-175)$ .}
    \label{fig:nonp_2D_fict_field}
\end{figure}

The light-induced  magnetic field formed by the optical tweezers has a component along the propagation axis, i.e. the $y$-axis, as well as components along the $z-$ and $x-$axes, due to tight focusing. The vector plot of the fictitious magnetic field from the Gaussian mode tweezers beam in the $xy-$plane is shown in Fig.~\ref{fig:vector field}(b) and the intensity distribution is shown in Fig.~\ref{fig:setup}(b). The 2D-plots of the $y-$component and the $x(z)-$component of the fictitious magnetic field are shown in Fig.~\ref{fig:nonp_2D_fict_field} (a,b), respectively. Note that, optical tweezers-based atom traps with a circularly polarized electric field were previously analyzed in~\cite{corwin1999spin}.

For optical tweezers, one can also use a Laguerre-Gaussian (LG) mode or a Hermite-Gaussian (HG) mode to modify the electric field profile allowing the high intensity region to be brought closer to the fiber surface. The electric field of a donut mode beam, LG$_{01}$ ($p=0, l=1$),  in the non-paraxial regime near the focal plane can be written as~\cite{youngworth2000focusing,winnerl2012universal,bhowmik2018tuning}

\begin{equation}
    \begin{split}
        &\boldsymbol{E_r} = [A \int_0^\alpha E_0(\theta)\, \sin 2\theta\, J_1(k\rho \sin\theta)\, \exp(i k z \cos\theta) \, \mathrm{d}\theta] \boldsymbol{\hat{e_r}},
        \\
        &\boldsymbol{E_{\phi}} = [2iA \int_0^\alpha E_0(\theta)\, \sin \theta\, J_1(k\rho \sin\theta)\, \exp(i k z \cos\theta) \, \mathrm{d}\theta] \boldsymbol{\hat{e_{\phi}}},
        \\
        &\boldsymbol{E_y}= [2iA \int_0^{\alpha} E_0(\theta) \, \sin^2 \theta \, J_0\left(k\rho \sin \theta\right) \exp\left(ikz \cos \theta\right) d\theta]\boldsymbol{\hat{e_y}},
        \\
        &E_0(\theta) = \exp\left[-\beta_0^2 \left( \frac{\sin\theta}{\sin\alpha} \right)^2 \right]J_1\left(2\beta_0\frac{\sin{\theta}}{\sin{\alpha}}\right)\sqrt{\cos{\theta}},
    \end{split}
\end{equation}

\noindent where $\rho$ is the radial coordinate from the beam center and A is a normalization constant. In a tightly focused beam $E_{\phi}$ does not generate any longitudinal electric field and propagates as a purely transverse polarization due to Maxwell equations~\cite{youngworth2000focusing}. The intensity profile of the LG$_{01}$  beam in the $xz-$plane is shown in Fig.~\ref{fig:setup}(c). The light-induced  magnetic field from the LG$_{01}$ beam can be expressed via the following equations

\begin{equation}\label{eq:fict_field}
    \begin{split}
    &\boldsymbol{B}_\mathrm{fict}^{\mathrm{tw}_y}=2\eta I_0^{\mathrm{LG}}\frac{\alpha^{\mathrm{v}}_{nJF}}{4\mu_B g_{nJF} F}\mathrm{Im}(\boldsymbol{E_r}\cdot\boldsymbol{E_{\phi}^*})\boldsymbol{\hat{e_y}},
    \\
    &\boldsymbol{B}_\mathrm{fict}^{\mathrm{tw}_{\phi}}=2\eta I_0^{\mathrm{LG}}\frac{\alpha^{\mathrm{v}}_{nJF}}{4\mu_B g_{nJF} F}[\mathrm{Im}(\boldsymbol{E_{r}}\cdot\boldsymbol{E_{y}^*})]\boldsymbol{\hat{e_{\phi}}},
    \\
    &\boldsymbol{B}_\mathrm{fict}^{\mathrm{tw}_r}=2\eta I_0^{\mathrm{LG}}\frac{\alpha^{\mathrm{v}}_{nJF}}{4\mu_B g_{nJF} F}[\mathrm{Im}(\boldsymbol{E_{y}}\cdot\boldsymbol{E_{\phi}^*})]\boldsymbol{\hat{e_r}},
    \end{split}
\end{equation}

\noindent where the peak intensity for the Laguerre-Gaussian mode $I_0^{\mathrm{LG}} \approx I_0^\mathrm{G}/\sqrt{2\pi}=\frac{2P}{\pi\omega_0^2\sqrt{2\pi}}$~\cite{allen1992orbital}.  The vector plot of the fictitious magnetic field from the LG$_{01}$ beam in the $xy-$plane is shown in Fig.~\ref{fig:vector field}(c).

When an atom is introduced into a combination of different fictitious magnetic fields, or a combination of a fictitious and a real magnetic field, the effective magnetic field can be calculated as a vector sum~\cite{zielonkowski1998optically, yang2008optically} with the resultant field defining the quantization axis, such that $\boldsymbol{B}_{\mathrm{eff}}=\boldsymbol{B}_{\mathrm{fict}}+\boldsymbol{B}_\mathrm{fict}^\mathrm{tw}+\boldsymbol{B}_\mathrm{bias}$, as demonstrated in experiments~\cite{zielonkowski1998optically,park2002optical}. From  Fig.~\ref{fig:vector field}, it can be seen that the fictitious magnetic fields produced by the hybrid OPTON platform have opposite directions on the positive $x$ side of the fiber.  Therefore, the effective magnetic field, $\boldsymbol{B}_{\mathrm{eff}}$, has a minimum amplitude at some distance from the ONF surface. The magnetic potential an atom experiences is 
\begin{equation}
	U_{\mathrm{mag}}=-\boldsymbol{\mu}\cdot\textbf{B}_{\mathrm{eff}},
\end{equation}
where $\boldsymbol{\mu}$ is the magnetic moment of the atom~\cite{KienEPJD2013}. Here, the potential  is formed for low-field seeking atoms. We set the quantization axis along the $z$-axis by adding a bias magnetic field of 3~G. Hence, the magnetic moment of the atom remains anti-parallel to the local effective magnetic field during atomic motion in the trap and the trapping potential can be simplified to

\begin{equation}\label{eq:8}
	U_{\mathrm{mag}}=\mu_{\mathrm{B}} g_{nJF}m_F|\textbf{B}_{\mathrm{eff}}|.
\end{equation}

\section{Results}

We consider a $^{87}$Rb atom in the $\mathrm{\left|5S_{1/2}, F=2, m_F=2\right>}$ ground state, which, for the fiber-guided mode with a free-space wavelength of $\lambda_{\mathrm{ONF}}=790.2$~nm, has a vector polarizability $\alpha_{nJ}^{\mathrm{v}}=55\times 10^{-5}~\mathrm{Hz\cdot m^2/V^2}$ and a scalar polarizability $\alpha_{nJF}^{\mathrm{sc}}\approx0~\mathrm{Hz\cdot m^2/V^2}$,  calculated using the Alkali-Rydberg-Calculator (ARC)~\cite{robertson2021arc}. We then find $\alpha^{\mathrm{v}}_{nJF}$ via $\alpha_{nJ}^{\mathrm{v}}$ through the following equation~\cite{KienEPJD2013}

\begin{equation}\label{eq:13}
	\begin{split}   
		\alpha^{\mathrm{v}}_{nJF}=&(-1)^{J+I+F+1}\sqrt{\frac{2F(2F+1)(J+1)(2J+1)}{2J(F+1)}} \\
		&\times\begin{Bmatrix}
			F&1&F\\
			J&I&J
		\end{Bmatrix}\alpha^{\mathrm{v}}_{nJ}
	\end{split}    
\end{equation}
\noindent to be $\sim 42\times 10^{-5}~\mathrm{Hz\cdot m^2/V^2}$.

The intensity of the tweezers light-induced magnetic field decays along the $z$-axis from the trapped atom position, see Fig.~\ref{fig:setup}(b,c), while the ONF guided mode field is homogeneous in intensity along the $z$-axis. Hence, there is no trapping potential created along the $z$-axis, see Appendix~\ref{app_A}. To overcome this issue, we introduce a detuning to the ONF guided mode wavelength, so that an attractive force for the ground state atoms is generated. We set $\lambda_\mathrm{ONF}$ to 787.9~nm instead of 790.2~nm, hence the scalar polarizability, $\alpha^{\mathrm{sc}}_{nJF}=12\times 10^{-5}~\mathrm{Hz\cdot m^2/V^2}$. The non-zero scalar polarizability results in an attractive potential, $U_{\mathrm{sc}}$. We also include atom-surface interactions using a van der Waals potential, which is non-negligible at distances less than $\sim$100~nm from the ONF surface~\cite{minogin2010manifestation}. This is introduced by $U_{\mathrm{vdW}}=-C_3/(r-r_{f})^3$, where $C_3=3.362\times 10^{-23}$~mK$\cdot\mathrm{m}^3$ for Rb~\cite{daly2014nanostructured} and $r_{\mathrm{f}}=175$~nm is the ONF radius. Hence, the total potential the atom experiences can be written as

\begin{equation}\label{eq:total U}
	\begin{split}
		U_{\mathrm{tot}}&=U_{\mathrm{mag}}+U_{\mathrm{sc}}+U_{\mathrm{vdW}}\\
		&=\mu_{\mathrm{B}} g_{nJF}m_F|\textbf{B}_{\mathrm{eff}}|-\frac{1}{4} \alpha^{\mathrm{sc}}_{nJF} |\boldsymbol{\mathcal{E}}_{\mathrm{ONF}}|^2-\frac{C_3}{(r-r_{\mathrm{f}})^3}
	\end{split}    
\end{equation}
\noindent where $|\boldsymbol{\mathcal{E}}_{\mathrm{ONF}}|^2$ is the square of the absolute value of the ONF guided mode electric field amplitude.

We set the polarization of the ONF guided mode to quasi-linear to increase the intensity of the evanescent field. This increases the amplitude of the fictitious magnetic field at the overlap with the tweezers beam. In addition, using a quasi-linear polarization creates confinement in the azimuthal direction unlike the quasi-circularly polarized case. The quasi-circular field is homogeneous along the azimuthal direction and therefore does not create a trapping potential along $\phi$, see Appendix~\ref{app_A}.

In addition, we calculate the potential, $U_{\mathrm{tot}}$, for  $\lambda_{\mathrm{ONF}}=\lambda_\mathrm{tw}=762$~nm, for which the scalar polarizability is four times larger than the vector polarizability. Therefore, the fictitious magnetic field part contribution to the potential  is negligible. We can then consider that the potential is constructed fully by the scalar AC Stark shift. We also calculate the potential, $U_{\mathrm{tot}}$, in the case of an attractive AC scalar shift for $\lambda_{\mathrm{ONF}}=\lambda_\mathrm{tw}=1064$~nm. We show that, in both cases, there is no trap formed for ground state $^{87}$Rb atoms, see Appendix~\ref{app_A}.

\subsection{Gaussian tweezers}

\begin{figure*}
	\centering
	\includegraphics[width=\linewidth]{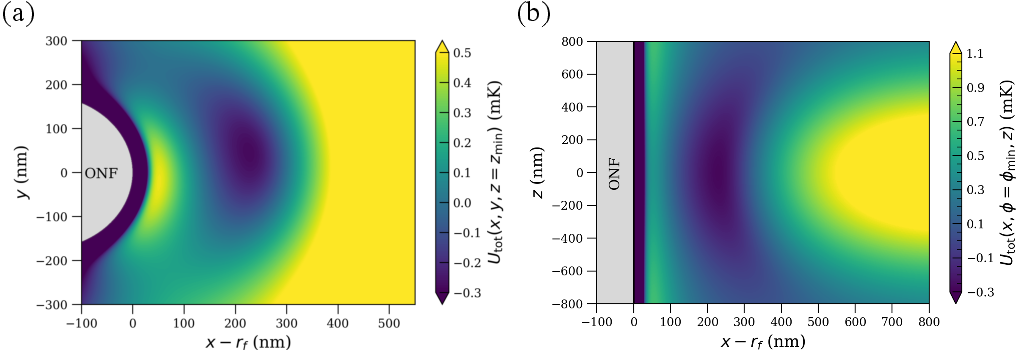}
	\caption{Two-dimensional plots of the light-induced, fictitious magnetic field trapping potential for a $^{87}$Rb atom in the $\mathrm{\left|5S_{1/2}, F=2, m_F=2\right>}$ ground state formed by $P_{\mathrm{ONF}}=1$~mW and $P_{\mathrm{tw}}=0.3$~mW: (a) $U_\mathrm{tot}(x,y,z=z_\mathrm{min})$  and (b) $U_\mathrm{tot}(x,\phi=\phi_\mathrm{min},z)$. The potential minimum is formed $\sim 220$~nm from the fiber surface  and the trap depth is $\sim 0.3$~mK. The small tilt in the trap position from the $y=0$ line is due to the $x$-component of the fictitious magnetic field, see Fig~\ref{fig:vector field}. Here, $\phi=\phi_\mathrm{min}$ and $z=z_\mathrm{min}$ are the coordinates of the trap minimum. The free-space wavelengths are $\lambda_\mathrm{ONF}=787.9$~nm and $\lambda_\mathrm{tw}=790.2$~nm for the fiber-guided mode and tweezers mode, respectively. The fiber radius is $r_f=175$~nm. Each plot has an offset so that the lowest point represents the trap depth, defined as the minimum potential barrier of the three-dimensional potential, $U_{\mathrm{tot}}(x,y,z)$. Unless otherwise stated, all configuration parameters, except the optical powers, are kept constant in subsequent figures to allow direct comparison between trapping configurations.}
	\label{fig:2D_traps_Gauss}
\end{figure*}

\begin{figure*}
	\centering
	\includegraphics[width=\linewidth]{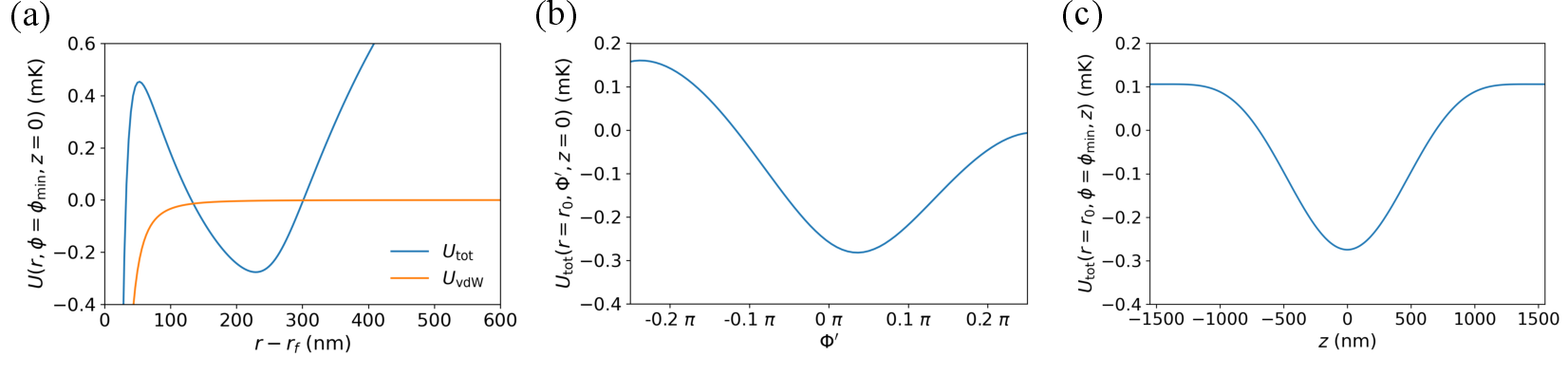}
	\caption{Calculated one-dimensional plots of the total trapping potential for a $^{87}$Rb atom in the $\mathrm{\left|5S_{1/2}, F=2, m_F=2\right>}$ ground state formed by $P_{\mathrm{ONF}}=1$~mW and $P_{\mathrm{tw}}=0.3$~mW in the (a) radial, $U_\mathrm{tot}(x,y=0,z=0)$, (b) azimuthal, $U_\mathrm{tot}(r=r_0(\Phi'),\Phi',z=0)$, and (c) longitudinal, $\mathrm{U_{tot}(r=r_0(z),y=0,z)}$, directions . Here, $r_0(\Phi')$ and $r_0(z)$ are the distances from the fiber surface to the trap minimum for fixed values of $\Phi'$ and $z$ respectively, see Fig.~\ref{fig:2D_traps_Gauss}(a,b). $\Phi'$ is the adjusted azimuthal component along the trapping potential in the $xy-$plane at the trap minimum.}
	\label{fig:1D_plots}
\end{figure*}

We calculate the total trapping potential for the $\left|5\mathrm{S}_{1/2}, F=2, m_F=2\right>$ state with the power in the Gaussian tweezers set to $0.3$~mW and the power in the ONF QL mode set to $1$~mW. The center of the tweezers beam is placed at a distance of $825$~nm from the ONF surface, with a waist, $w_0=0.5~\mu$m. The QL polarization of the ONF guided mode is set along the $x$-axis. The 2D plots of the trapping potential in the $xy-$ and $xz-$planes are shown in Fig.~\ref{fig:2D_traps_Gauss}(a,b).   The minimum of the trapping potential is produced approximately 200~nm from the surface of the ONF with a potential depth, $U_\mathrm{tot}\approx 0.8$~mK, as shown in Fig.~\ref{fig:1D_plots}(a). We also show that the produced potential traps atoms in both the azimuthal ($\Phi'$) and longitudinal ($z$) directions on 1-dimensional plots, see Fig.~\ref{fig:1D_plots}(b,c).

One concern is that some of the tweezers light may scatter from the ONF surface and locally perturb the electric field. We calculate the ratio of the intensity of the light from the tweezers  near the ONF surface to the intensity at the trap minimum to be around 5$\%$. In addition, the reflected light does not necessarily interfere with the tweezers light in the trap area, therefore, we neglect the perturbation of the intensity profile from the reflection. However, further investigation may be needed to better understand the effects of the reflected fields.

\begin{figure}
	\centering
	\includegraphics[width=\linewidth]{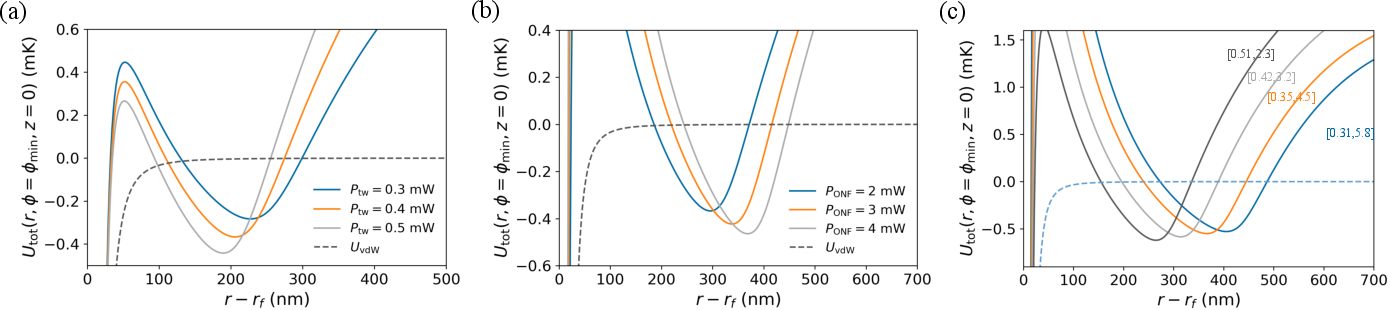}
	\caption{Van der Waals potential (dotted lines) and radial trapping potential (solid lines) for a $^{87}$Rb atom in the $\mathrm{\left|5S_{1/2}, F=2, m_F=2\right>}$ ground state formed by (a) $P_{\mathrm{ONF}}=1.0$~mW  and $P_{\mathrm{tw}} =0.3,~0.4$, and 0.5~mW, (b) $P_{\mathrm{tw}}=0.3$~mW and $P_{\mathrm{ONF}}= 2.0,~3.0$, and 4.0~mW, and (c) [$P_\mathrm{tw}\mathrm{(mW)}, P_\mathrm{ONF}\mathrm{(mW)}]=$[0.31, 5.8] (blue), [0.35, 4.5] (orange), [0.42, 3.2] (gray), and [0.51,2.3] (black). }
	\label{fig:diff_powers}
\end{figure}

The trap depth and minimum position strongly depend on both the power of the tweezers beam and the power of the ONF guided light. To illustrate this, we vary the power of the tweezers beam, $P_{\mathrm{tw}}$, from $0.3$~mW to $0.5$~mW in steps of $0.1$~mW while keeping the power of the ONF guided mode, $P_{\mathrm{ONF}}$, constant at 1~mW, see Fig.~\ref{fig:diff_powers}(a). It can be seen that, by changing the power in the tweezers, one can easily move the trap minimum from around $230$~nm to $190$~nm, while increasing  the trap depth from $0.3$~mK to $0.45$~mK. To produce a trapping potential with a similar trap depth using a two-color fiber-based dipole trap~\cite{vetsch2010optical} one would need three times more power in the ONF guided mode.

\begin{table}
\caption{\label{table: gauss}Parameters for the trapping configurations shown in Fig.~\ref{fig:diff_powers}.  $P_{\mathrm{ONF}}$ is the power of the fiber-guided mode, $P_{\mathrm{tw}}$ is the tweezers power, $r_0$ is the trap distance from the ONF surface, $U_0$ is the trap depth, $\omega_r$, $\omega_{\Phi'}$ and $\omega_z$ are the radial, azimuthal and longitudinal trapping frequencies, respectively, $R_\mathrm{sc}$ is the off-resonant scattering rate, and $\eta$ is the Lamb-Dicke parameter. All values are as used in Fig.~\ref{fig:diff_powers}.}

\begin{tabular}{
    @{}
    >{\columncolor{white}}c
    >{\columncolor{gray!10}}c|
    >{\columncolor{white}}c
    >{\columncolor{gray!10}}c
    >{\columncolor{white}}c
    >{\columncolor{gray!10}}c
    >{\columncolor{white}}c
    >{\columncolor{gray!10}}c
    >{\columncolor{white}}c
}

$P_{\mathrm{ONF}}$& $P_{\mathrm{tw}}$ & $r_0$ & $U_0$ & $\omega_r/2\pi$ & $\omega_{\Phi'}/2\pi$ & $\omega_z/2\pi$ & $R_\mathrm{sc}$ & $\eta$\\
(mW) & (mW) & (nm) & (mK) & (MHz) & (kHz) & (kHz) & (1/s) & \\
\hline
\multicolumn{8}{c}{ Gaussian mode optical tweezers} \\
\hline
1 & 0.3 & 227 & 0.28 & 0.35 & 112 &  44 & 35.2 & 0.29 \\  
		
1 & 0.4 & 205 & 0.37 & 0.41 & 132 & 48 & 42.4 & 0.28 \\
		
1 & 0.5 & 189 & 0.44 & 0.48 & 150 & 52 & 48.9  & 0.27\\
		
\hline
2 &  0.3 & 296 & 0.37 & 0.38 & 110 &  50 & 42.5 & 0.28\\
		
3 & 0.3 & 338 & 0.42 & 0.39 & 109 &  54 & 47.2 & 0.27 \\
		
4 & 0.3 & 369 & 0.46 & 0.40 & 108 &  57 & 50.8  &  0.26\\

\end{tabular}

\end{table}

Similarly, we calculate the trapping potentials when keeping the power in the tweezers, $P_{\mathrm{tw}}$, constant at $0.3$~mW and varying the power in the ONF guided mode, $P_\mathrm{ONF}$, from $2$~mW to $4$~mW, see Fig.~\ref{fig:diff_powers}(b). If the effective magnetic field reaches low values at the trap minimum, high rates of spin flipping, i.e., changes of the $m_F$ state from positive to negative, can occur.  This would be detrimental to the trap lifetime - once the sign of the $m_F$ state changes the atom is repelled from the trap. However, the addition of a quantization magnetic field, which we set to an amplitude of $B_\mathrm{bias}=3$~G along the $z$-axis, prevents the loss of atoms from the trap and keeps the spin flip rates, $\Gamma_{\mathrm{sf}}=\frac{\pi \omega_r}{2} exp\left(-\frac{\pi\mu_B g_{nJF}|\textbf{B}|}{2\hbar\omega_r}\right)$~\cite{sukumar1997spin}, on the order of $10^{-4}$~s$^{-1}$. 

For each configuration we compute the trap depth, $U_0$, and trap minimum position relative to the ONF surface, $r_0$, the radial, azimuthal and longitudinal trap frequencies, $\omega_r$, $\omega_{\Phi'}$ and $\omega_z$ respectively, where $\omega_i=\sqrt{\left. \frac{1}{m}\frac{\partial^2 U}{\partial x_i^2}\right|_{x_{\mathrm{min}}}}$~\cite{grimm2000optical}, and the off-resonant scattering rate for both the tweezers and the ONF electric fields, $R_{\mathrm{sc}}$, see Table~\ref{table: gauss}. The upper limit of the atom lifetime in the trap is set by recoil heating~\cite{goban2012demonstration}; however, Raman scattering can contribute to an additional loss. Due to off-resonant Raman scattering, an atom may undergo a change in the $m_F$ state that affects the trap potential. Raman scattering gives the worst case estimate of the atomic lifetime in the trap, $\tau_\mathrm{R}\sim1/R_\mathrm{sc}$. The coherence time is also limited by scattering processes, the dominant one being Raman scattering, since the trap wavelength lies between the $^{87}$Rb $D_1$ and $D_2$ lines. In our configuration, typical values of $R_\mathrm{sc}$ are on the order of $40$~s$^{-1}$, setting the worst-case trap lifetime limit, $\tau_\mathrm{trap}$ to $\sim25$~ms. The off-resonant scattering rate can be calculated from~\cite{Bai_2020}

\begin{equation}
    R_\mathrm{sc}=\frac{\Gamma}{2}\frac{I/I_\mathrm{sat}}{1+4(\Delta/\Gamma)^2+(I/I_\mathrm{sat})},
\end{equation}
\noindent where $I_\mathrm{sat}$ is the average saturation intensity for all possible transitions from a given $m_F$ state, $I$ is the optical field intensity, $\Delta$ is the detuning from the nearby transition, and $\Gamma$ is the spontaneous decay rate. In addition, a typical value of the Lamb-Dicke parameter $\eta=\frac{2\pi}{\lambda}\sqrt{\frac{\hbar}{2m\omega}}$ for our trap potentials, see Table~\ref{table: gauss}, is on the order of 0.3, which shows that we are in the upper limit of the Lamb-Dicke regime~\cite{roghani2008trapped}. Therefore, there is a low, albeit non-vanishing, probability of the atomic internal motional states coupling.

In addition, we show that by controlling the powers of the optical tweezers and the ONF guided mode in a particular way one can keep the trap depth identical while moving the trap position from the ONF surface, see Fig.~\ref{fig:diff_powers}(c).

\subsection{Laguerre-Gaussian tweezers}

\begin{figure*}
	\centering
	\includegraphics[width=\linewidth]{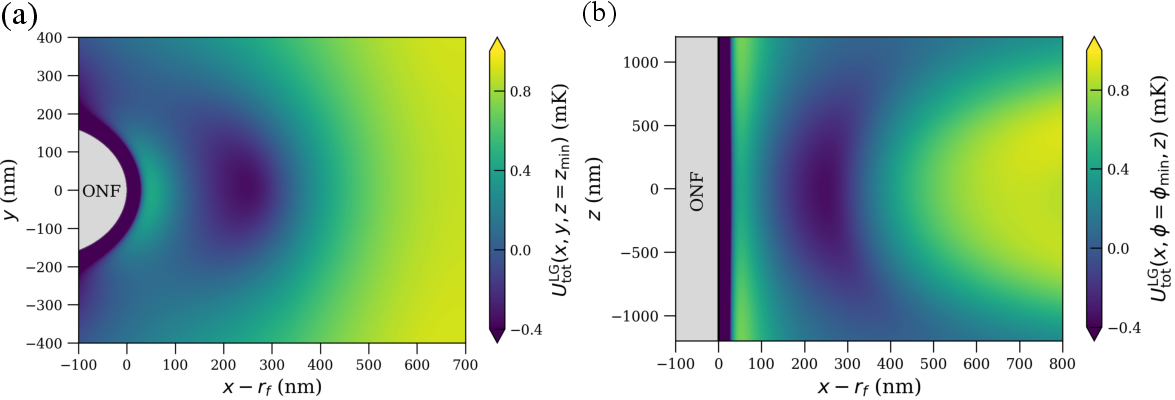}
	\caption{Two-dimensional plots of the light-induced, fictitious magnetic field trapping potential for a $^{87}$Rb atom in the $\mathrm{\left|5S_{1/2}, F=2, m_F=2\right>}$ ground state formed by $P_{\mathrm{ONF}}=1$~mW and $P_{\mathrm{LG}}=0.3$~mW for (a) $U_\mathrm{tot}(x,y,z=z_\mathrm{min})$ and (b) $U_\mathrm{tot}(x,\phi=\phi_\mathrm{min},z)$. Here, $\phi=\phi_\mathrm{min}$ and $z=z_\mathrm{min}$ are the coordinates of the trap minimum. The potential minimum is formed $\sim250$~nm from the fiber surface and the trap depth is $\sim 0.4$~mK.}
	\label{fig:2D_plots_LG}
\end{figure*}

\begin{figure*}
	\centering
	\includegraphics[width=\linewidth]{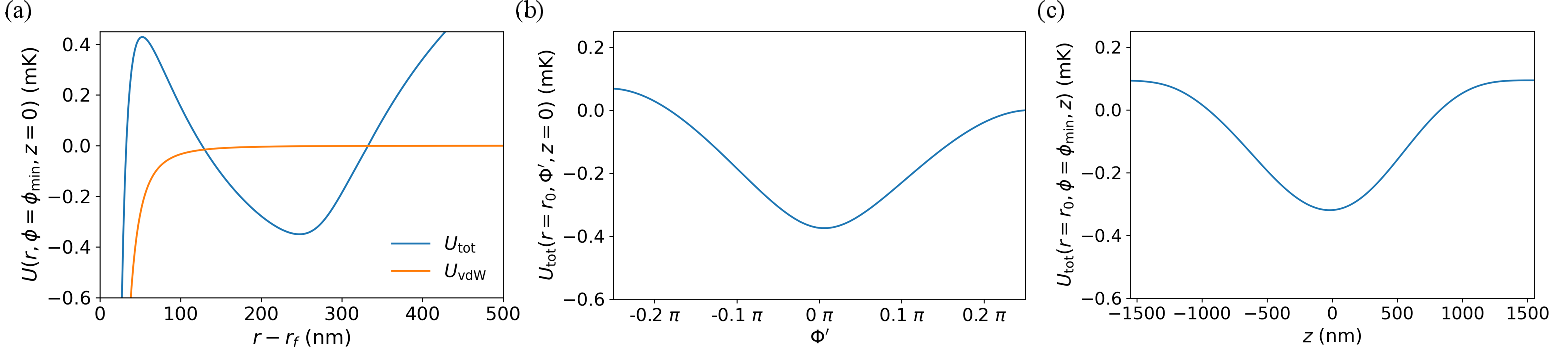}
	\caption{One-dimensional plots for a $^{87}$Rb atom in the $\mathrm{\left|5S_{1/2}, F=2, m_F=2\right>}$ ground state formed by $P_{\mathrm{ONF}}=1$~mW and $P_{\mathrm{LG}}=0.3$~mW. (a) Van der Waals potential (orange), $U_{\mathrm{vdW}}$, and the total trapping potential (blue), $\mathrm{U_{tot}(r,\phi=\phi_\mathrm{min},z=z_\mathrm{min})}$, in the radial direction. The light-induced magnetic field trapping potential in (b) the azimuthal $U_\mathrm{tot}(r=r_0,\Phi')$ and (c) the longitudinal, $U_\mathrm{tot}(r=r_0,z)$ directions. Here, $\Phi'$ is the adjusted azimuthal component along the trapping potential in the $xy-$plane at the trap minimum $(r=r_0, z=z_\mathrm{min})$.}
	\label{fig:1D_LG}
\end{figure*}

In the same manner, we calculate the total trapping potential for the $\left|5\mathrm{S}_{1/2}, F=2, m_F=2\right>$ state with the power of the circularly polarized LG$_{01}$ ($p=0,~l=1)$ tweezers beam, $P_{LG}$, set to $1$~mW and the power of the ONF QL mode set to $1$~mW. The center of the tweezers beam is positioned 1325~nm away from the ONF surface, with a beam waist of $0.5~\mu$m. The QL polarization of the ONF-guided mode is aligned along the $x$-axis. Figure~\ref{fig:2D_plots_LG}(a,b) shows 2D plots of the trapping potential in the $xy-$ and $xz-$planes, respectively. From the radial 1-dimensional plot shown in Fig.~\ref{fig:1D_LG}(a), one can see that the trapping potential reaches a minimum approximately 250~nm from the ONF surface, with a depth of around 0.4~mK. The 1D trapping potential along the azimuthal ($\Phi'$) and longitudinal ($z$) directions are also shown in Fig.~\ref{fig:1D_LG}(b,c), respectively.

\begin{figure}
	\centering
	\includegraphics[width=\linewidth]{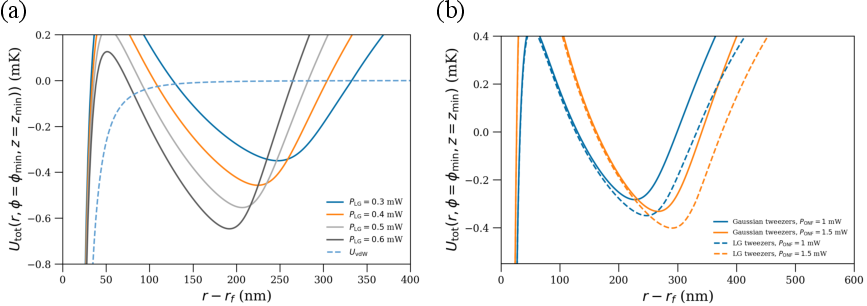}
    
	\caption{(a) Van der Waals potential (dotted line) and radial trap potentials (solid lines) for $P_\mathrm{ONF}=1$~mW and $P_\mathrm{LG}=0.3,~0.4,~0.5$ and $0.6$~mW.  (b) Radial trap potentials for the $\left|5\mathrm{S}_{1/2}, F=2, m_F=2\right>$ ground state formed by $P_\mathrm{LG}=0.3$~mW (dashed lines) and $P_\mathrm{tw}=0.3$~mW (solid lines) and $P_\mathrm{ONF}=1$~mW (blue) and $P_\mathrm{ONF}=1.5$~mW (orange). All plots are for  a $^{87}$Rb atom in the $\mathrm{\left|5S_{1/2}, F=2, m_F=2\right>}$ ground state.}
	\label{fig:LG_diff_powers}
\end{figure}

In addition, we look at how the trap depth and the trap minimum position change depending on the power of the LG$_{01}$ mode tweezers beam while keeping the power of the ONF guided mode, $P_{\mathrm{ONF}}$, constant at $1$~mW. To illustrate this, we plot the trapping potential in the radial direction for the power of the tweezers beam, $P_{\mathrm{LG}}$, set to $0.3$~mW, $0.4$~mw, $0.5$~mW and $0.6$~mW, see Fig.~\ref{fig:LG_diff_powers}(a). The trap depth increases from $\sim 0.4$~mK to $\sim0.7$~mK and the trap minimum moves from $\sim 250$~nm to $\sim 190$~nm from the surface of the ONF when  $P_{\mathrm{LG}}$ is changed  from $0.3$~mW to $0.6$~mW, see Fig.~\ref{fig:LG_diff_powers}(a). By further increasing the power of the tweezers, one can move the trap minimum to less than $100$~nm from the ONF surface, where the van der Waals potential becomes a limiting factor both for the trap depth and the minimum position.

Similarly, for each power of the LG$_{01}$ tweezers we compute the trap depth, $U_0$, and trap minimum position, $r_0$, the radial, azimuthal and longitudinal trap frequencies, $\omega_r$, $\omega_{\Phi'}$ and $\omega_z$ respectively, off-resonant scattering rate, $R_\mathrm{sc}$, Lamb-Dicke parameter, $\eta$, and display them in Table~\ref{table:params_LG}. With the same power in the ONF-guided mode, the LG tweezers produce $\approx 1.5$~times deeper potential than the Gaussian tweezers of the same power, i.e., $P_{\mathrm{LG}}=P_{\mathrm{tw}}$. This difference arises from the distinct electric field intensity distributions of the Gaussian and LG$_{01}$  tweezers beams.

\begin{table}[]
\caption{\label{table:params_LG}Parameters for the trapping configurations shown in Fig.~\ref{fig:LG_diff_powers}(a).  $P_{\mathrm{ONF}}$ is the the fiber-guided power, $P_{\mathrm{LG}}$ is the LG$_{01}$ mode tweezers power, $r_0$ is the trap distance from the ONF surface, $U_0$ is the trap depth, $\omega_r$, $\omega_{\Phi'}$ and $\omega_z$ are the radial, azimuthal and longitudinal trapping frequencies, respectively, $R_\mathrm{sc}$ is the off-resonant scattering rate, and $\eta$ is the Lamb-Dicke parameter. All values are as used in Fig.~\ref{fig:LG_diff_powers}(a).}

\begin{tabular}{
    >{\columncolor{white}}c
    >{\columncolor{gray!10}}c|
    >{\columncolor{white}}c
    >{\columncolor{gray!10}}c
    >{\columncolor{white}}c
    >{\columncolor{gray!10}}c
    >{\columncolor{white}}c
    >{\columncolor{gray!10}}c
    >{\columncolor{white}}c
}

$P_{\mathrm{ONF}}$& $P_{\mathrm{LG}}$ & $r_0$ & $U_0$ & $\omega_r/2\pi$ & $\omega_{\Phi'}/2\pi$ & $\omega_z/2\pi$ & $R_{\mathrm{sc}}$ & $\eta$ \\
(mW) & (mW) & (nm) & (mK) & (MHz) & (kHz) & (kHz) & (1/s) &  \\
\hline
\multicolumn{8}{c}{ Laguerre-Gaussian mode optical tweezers} \\
\hline
1 & 0.3 & 247 & 0.35 & 0.41 & 90 & 33 & 26 & 0.33\\  
		
1 & 0.4 & 225 & 0.46 & 0.52 & 76 & 39 &  31 & 0.31 \\
		
1 & 0.5 & 207 & 0.56 & 0.62 & 63 & 42 &  37 &  0.30\\
		
1 & 0.6 & 193 & 0.65 & 0.7 & 39 &  46 & 42 & 0.28 \\

\end{tabular}

\end{table}

We compare the performance of the two trap configurations created by the Gaussian tweezers beam and the LG$_{01}$ tweezers beam with the same powers, $P_{\mathrm{tw}}=P_{\mathrm{LG}}$, while changing the power of the ONF guided mode, see Fig.~\ref{fig:LG_diff_powers}(b). The LG mode offers a deeper trapping potential while maintaining comparable tunability of the trap minimum position and depth to that of a Gaussian mode when the power of the fiber-guided mode is varied.

\section{Conclusion}
In this work, we have presented a novel atom trapping scheme based on a nanofiber-optical tweezers configuration OPTON and provided a detailed analysis of key trap parameters for $^{87}$Rb atoms. This scheme could be feasible for other alkali atoms and, in general, for atoms with non-zero vector polarizability and an accessible tune-out wavelength for the scalar light shift. The proposed fictitious magnetic trap achieves depths sufficient for trapping laser-cooled atoms, using optical powers comparable to those required for conventional two-color nanofiber traps and optical tweezers. Our scheme provides several key advantages over the configuration where an optical tweezers illuminates an optical waveguide, creating a standing-wave dipole potential for trapping an atom. For one, OPTON allows for significant tunability of the trap minimum position relative to the waveguide surface, enabling adjustable coupling of light emitted by the trapped atoms into the fiber-guided mode. Additionally, it avoids the need for direct illumination of the ONF by the tweezers light, thereby diminishing the risk of ONF damage from heat dissipation caused by scattered light and excitation of nanofiber vibration modes~\cite{song2024measurement}. In our scheme, the atom position — or trap minimum — can be readily adjusted as the powers of both the ONF-guided mode and the optical tweezers are controlled via an acousto-optic modulator. When manipulating trapped atoms one should change the trap position in an adiabatic manner compared to the trap frequencies. The calculated trap frequencies and AOM rise times allow for the proposed scheme to make adjustments on timescales comparable to acousto-optic deflector-based optical tweezers.

We analyzed and compared the trap configurations for both Gaussian and Laguerre-Gaussian mode tweezers beams. We concluded that the Laguerre-Gaussian tweezers provides similar control over the trap depth, as well as the minimum trap position from the ONF surface in comparison to the Gaussian mode tweezers. Additionally, we demonstrated that the LG mode tweezers offers a deeper trap compared to the Gaussian mode at similar powers. This advantage arises from the different intensity distributions between the LG and Gaussian modes. 

In typical configurations of our proposed OPTON fictitious magnetic field trap, the trap minimum is positioned $100–400$~nm from the ONF surface, with trap depths ranging from $0.3$ to $0.8$~mK. A trap depth of more than $0.3$~mK enables the loading of atoms that have been laser-cooled using a magneto-optical trap. Atoms may be loaded adiabatically from a MOT into the fictitious magnetic field trap as done for conventional wire traps~\cite{fortagh1998miniaturized}. In a MOT and optical molasses, atoms occupy all $m_F$ states and only atoms with $m_F<0$ will be loaded. One could transfer the atoms into the $m_F=-2$ state via a Raman process either using a free-space beam or through the ONF~\cite{meng2018near}. Alternatively, atoms could be transferred into the OPTON configuration from a conventional two-color ONF trap~\cite{vetsch2010optical} by adiabatically ramping down the blue- and red-detuned fields while simultaneously ramping up the ONF-guided and optical tweezers fields of the OPTON scheme, along with a gradual increase of the bias magnetic field. Another approach to load atoms into the OPTON trap would be to adiabatically transfer them from a conventional optical tweezers operating at $\lambda=1064$~nm into an overlapping OPTON tweezers at $\lambda=790.2$~nm, while gradually varying their optical powers. During this process, the bias magnetic field and the ONF-guided field could remain constant.

Our proposed trapping scheme combines fictitious magnetic fields from the tweezers and ONF-guided light, and a bias magnetic field. Therefore, the stability of the field sources is important. Fluctuations in the fictitious magnetic fields are determined by the laser stability, which is typically on the order of $0.1 \%$ RMS~(root-mean-squared). In contrast, fluctuations in the bias magnetic field, when produced by a pair of Helmholtz coils, are determined by the current noise in the coils and are estimated to be about 1 mG, which is about $10^{-4}$ of the suggested bias field value of 3~G. Both fluctuation levels are sufficiently low to ensure stable trapping. Another factor relevant to the practical implementation of the trapping scheme is the vibration of the optical nanofiber itself. For a vertically suspended nanofiber of 5 cm length at room temperature, the estimated frequency of the fundamental vibration mode is a few Hz. It is much slower than the timescale for any operation with cold atoms, therefore it is not expected to impose any limitations [32]. However, further investigation of the excitation and damping of ONF vibrational modes in vacuum is merited.

\section*{Acknowledgments}
The authors would like to thank the Scientific Computing and Data Analysis Section at the Okinawa Institute of Science and Technology Graduate University (OIST). This work was supported by funding from OIST.  S.N.C. acknowledges support from the Japan Society for the Promotion of Science (JSPS) Grant-in-Aid No. 24K08289. 

\section*{Data Availability Statement}
The data is available from the corresponding authors upon reasonable request.

\appendix\section{}\label{app_A}

\begin{figure}[h]
    \centering
    \includegraphics[width=\linewidth]{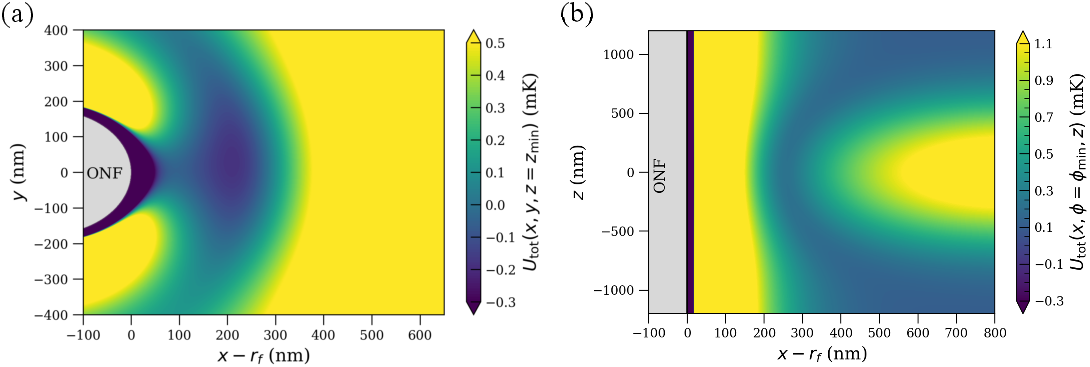}
    \caption{Two-dimensional plots of the light-induced, fictitious magnetic field trapping potential $\mathrm{U_{tot}(x,y,z)}$ for a $^{87}$Rb atom in the $\mathrm{\left|5S_{1/2}, F=2, m_F=2\right>}$ ground state formed by $P_{\mathrm{ONF}}=1$~mW and $P_{\mathrm{tw}}=0.3$~mW. (a) Use of a quasi-circularly polarized fiber-guided mode. The potential minimum in the radial and azimuthal directions is formed  $\sim 220$~nm from the fiber surface, however is only around $0.1$~mK. $\lambda_\mathrm{ONF}=787.9$~nm and $\lambda_\mathrm{tw}=790.2$~nm. (b) $\lambda_\mathrm{ONF}=\lambda_\mathrm{tw}=790.2$~nm. The radial potential minimum is formed  $\sim 250$~nm from the fiber surface. No trapping potential is formed in the longitudinal direction. }
    \label{fig:no_confinement}
\end{figure}

\begin{figure}[h]
    \centering
    \includegraphics[width=\linewidth]{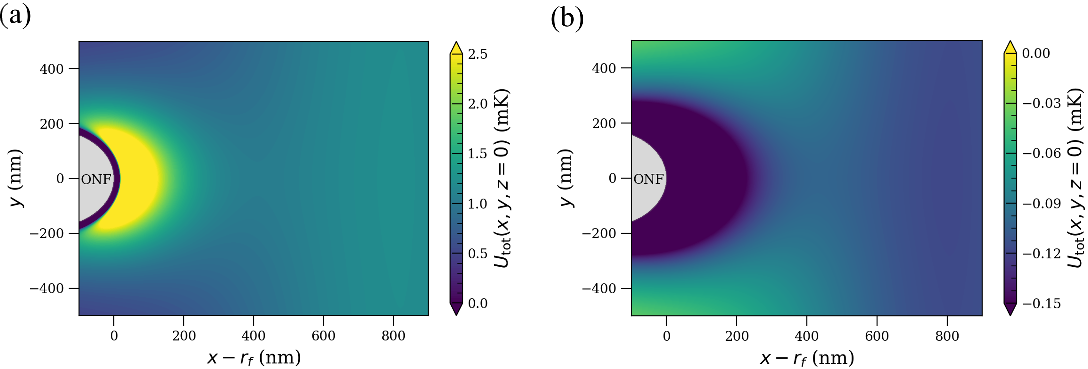}
    \caption{Two-dimensional plots of the potential $\mathrm{U_{tot}(x,y,z=0)}$ for a $^{87}$Rb atom in the $\mathrm{\left|5S_{1/2}, F=2, m_F=2\right>}$ ground state formed by $P_{\mathrm{ONF}}=1$~mW and $P_{\mathrm{tw}}=0.3$~mW for (a) $\lambda_\mathrm{ONF}=\lambda_\mathrm{tw}=762$~nm and (b) $\lambda_\mathrm{ONF}=\lambda_\mathrm{tw}=1064$~nm.}
    \label{fig:scalar_only}
\end{figure}

In the main text we justified the choice of using a quasi-linearly polarized guided mode in the ONF. In this section we show that the quasi-circularly polarized guided mode does not create a deep trapping potential. We calculate the trapping potential, $U_{\mathrm{tot}}$, for a quasi-circularly polarized guided mode in the ONF with a power of 1~mW and a Gaussian optical tweezers with a power of 0.3~mW. The free-space wavelengths $\lambda_\mathrm{ONF}=787.9$~nm and $\lambda_\mathrm{tw}=790.2$~nm are chosen for the fiber-guided mode and the tweezers mode, respectively. As one can see in  Fig.~\ref{fig:no_confinement}(a), a trapping potential of around $0.1~$mK depth is created in the azimuthal and the radial directions. However, the trap depth of $0.1~$mK is suboptimal and does not ensure stable and long trapping of cold atoms.

We also calculate the trapping potential, $U_{\mathrm{tot}}$, with zero wavelength detuning of the ONF-guided mode. We use a QL polarized guided mode in the ONF with a power of 1~mW and a Gaussian optical tweezers with a power of 0.3~mW. The free-space wavelength is chosen to be $\lambda_\mathrm{ONF}=\lambda_\mathrm{tw}=790.2$~nm for both the fiber-guided mode and the tweezers mode. As one can see in  Fig.~\ref{fig:no_confinement}(b), no trapping barrier is created in the longitudinal direction.

Moreover, we show that the potential, $U_\mathrm{tot}(x,y,z)$, cannot be used to trap atoms when only the scalar part of the AC Stark shift is present. Two-dimensional plots of the $U_\mathrm{tot}(x,y,z)$ at $z=0$ in the case of $\lambda_\mathrm{ONF}=\lambda_\mathrm{tw}=762$~nm and $\lambda_\mathrm{ONF}=\lambda_\mathrm{tw}=1064$~nm are shown in the Fig.~\ref{fig:scalar_only}(a,b) respectively.

\bibliography{refs}

\end{document}